\documentclass[11pt]{article}
\usepackage{mathtools,amssymb,amsfonts,amsthm,graphicx}
\usepackage[numbers,compress]{natbib}
\usepackage{fullpage}

\newcommand{\avg}[1]{\langle#1\rangle}
\newcommand{\cdf}{C(\lambda|\V x;t)}
\newcommand{\chis}{\chi_\mathsf S}
\newcommand{\dd}{{\rm d}}
\newcommand{\dy}[1]{\frac{\dd#1}{\dd y}}
\newcommand{\Enet}{E_{\rm net}}
\newcommand{\eprt}{\epsilon^{1/2}}
\newcommand{\Eq}[1]{(\ref{#1})}
\newcommand{\Eqs}[2]{(\ref{#1}) and (\ref{#2})}
\newcommand{\Fig}[1]{figure\,\ref{#1}}
\newcommand{\FF}[1]{Figure\,\ref{#1}}
\newcommand{\partialy}[1]{\frac{\partial#1}{\partial y}}
\newcommand{\partialgma}[1]{\frac{\partial#1}{\partial\gamma}}
\newcommand{\partialpp}[1]{\frac{\partial^2#1}{\partial\hat\psi^2}}
\newcommand{\partialt}[1]{\frac{\partial#1}{\partial t}}
\newcommand{\partialr}[1]{\frac{\partial#1}{\partial r}}
\newcommand{\qmax}{q_\mathrm{max}}
\newcommand{\qmin}{q_{\rm min}}
\newcommand{\Qvent}{Q_{\rm vent}}

\newcommand{\V}[1]{\vec{#1}}

\newcommand{\Rnet}{R_{\rm net}}
\newcommand{\sect}[1]{section\,\ref{#1}}
\newcommand{\Sect}[1]{Section\,\ref{#1}}
\newcommand{\step}{\mathcal H}

\def\erf{\operatorname{erf}\!}
\def\erfc{\operatorname{erfc}\!}

\let\ssection=\section
\renewcommand{\section}{\setcounter{equation}{0}\ssection}

\title{The effect of coherent stirring on\,the\,\mbox{advection--condensation} of water vapour}

\author{Yue-Kin Tsang\footnote{Corresponding author: \texttt{y.tsang@exeter.ac.uk}. Present address: Centre for Astrophysical and Geophysical Fluid Dynamics, University of Exeter, Exeter, EX4 4QF, UK.}\mbox{~~}and Jacques Vanneste \\ \normalsize{School of Mathematics and Maxwell Institute for Mathematical Sciences} \\ \normalsize{University of Edinburgh, Edinburgh, EH9 3FD, UK}}

\begin{document}

\maketitle

\begin{abstract}
Atmospheric water vapour is an essential ingredient of weather and climate. Key features of its distribution can be represented by kinematic models which treat it as a passive scalar advected by a prescribed flow and reacting through condensation. Condensation acts as a sink that maintains specific humidity below a prescribed, space-dependent saturation value. In order to investigate how the interplay between large-scale advection, small-scale turbulence and condensation controls the moisture distribution, we develop simple kinematic models  which combine a single circulating flow with a Brownian-motion representation of turbulence. We first study the drying mechanism of a water-vapour anomaly released inside a vortex at an initial time. Next, we consider a cellular flow with a moisture source at a boundary. The statistically steady state attained shows features reminiscent of the Hadley cell such as boundary layers, a region of intense precipitation and a relative humidity minimum. Explicit results provide a detailed characterisation of these features in the limit of strong flow.
\end{abstract}

\section{Introduction}
\label{intro}

Liquid water evaporates from land and ocean into the atmosphere. The interaction between the subsequent transport and condensation of this evaporated water gives rise to intriguing distributions of water vapour in the atmosphere: for example, persistent relative humidity minima are observed in the subtropics \cite{Sherwood10,OGorman11}, and bimodal distributions have been reported in the tropics \cite{Zhang03}. Knowledge of the full distribution of atmospheric humidity is crucial for understanding the Earth's energy balance and climate. This is because the absorption of outgoing long-wave radiation by water vapour increases nonlinearly (roughly logarithmically) with specific humidity \cite{Pierrehumbert10}. The atmospheric moisture distribution and transport is also closely linked to global and regional precipitation patterns which have high social and economic impacts \cite{metoff14}.

A framework to explain key features of the atmospheric humidity distribution is the advection--condensation model \cite{Pierrehumbert07,Sherwood10}. In this model, a moist air parcel is transported through the atmosphere's saturation humidity field and condensation occurs when its humidity exceeds the local saturation value. The excessive water is rained out of the system. As a result, the humidity at a particular location is equal to the minimum saturation value the air parcel has encountered since leaving the moisture source. Critically, all complex cloud-scale microphysics and molecular diffusion are excluded from this model \cite{Sherwood10}. Research over the last several decades has demonstrated the value of the idea of advection--condensation. Brewer in 1949 was able to deduce the existence of a general circulation in the stratosphere from water vapour distribution measurement \cite{Brewer49}. More recently, many studies have reconstructed humidity field in the troposphere \cite{Yang94, Sherwood96, Salathe97, Pierrehumbert98b, Brogniez09} and the stratosphere \cite{Liu10} by simulating particle trajectories using observed wind fields.

The success in numerical and observational studies has led to theoretical investigations of the advection--condensation model in idealised settings. A continuum formulation of the model, with the water vapour distribution represented by a coarse-grained field \cite{OGorman06,OGorman11}, is prone to produce overly saturated air \cite{Pierrehumbert07}. Here, we employ a Lagrangian particle formulation. A few previous works have taken this approach, starting with Pierrehumbert et al. \cite{Pierrehumbert07} who considered an ensemble of moist air parcels undergoing Brownian motion and condensation in one dimension. Among other results, they obtained analytically the time-dependent probability distribution function (PDF) of the local specific humidity when  initially saturated parcels are allowed to dry in the absence of moisture source---stochastic drying. The stochastic drying problem where the parcel velocity has a finite correlation time was solved by O'Gorman and Schneider \cite{OGorman06}. Sukhatme and Young \cite{Sukhatme11} studied Brownian parcels forced by a moisture source located at one end of a bounded one-dimensional domain and derived an exact solution for the water-vapour PDF of the resulting statistically steady state. A generalisation of this steady-state problem to the case of time-correlated parcel velocity was considered by Beucler \cite{Beucler16}. All these studies employ a one-dimensional Lagrangian velocity with no spatial correlation to mimic turbulent motions. However, analysis of observational data \cite{Trenberth03,Schneider06} and idealised simulations \cite{Boutle11} demonstrate that synoptic-scale eddies play an important role in atmospheric transport. Pauluis et al. \cite{Pauluis10} have also shown that the global moisture circulation can be viewed as a single overturning cell in moist isentropic coordinates. Roughly speaking, water vapour evaporated into the planetary boundary layer is drawn toward the tropics where it is transported upward. Large-scale advection then carries the moisture from the tropical upper troposphere to other regions where the air subsides \cite{Sherwood10}.

In this paper, we aim to gain insight into the effects of coherent stirring on the transport and distribution of water vapour. We consider a two-dimensional advection--condensation system where the velocity of an air parcel consists of a large-scale circulation and a small-scale stochastic component. We use this idealised model to investigate how the large- and small-scale velocities interact to produce the resulting humidity distribution and answer questions such as: How does the large-scale circulation create an area of low relative humidity? How does the precipitation pattern change with the strength of the circulation?

Following the presentation of our model in \sect{sec:model}, we investigate in \sect{sec:drying} the drying of a moisture patch in the presence of a single vortex and no moisture source. The drying process consists of an initial fast advective stage and a later slow stochastic stage. In the limit of strong circulation, we obtain an analytical expression for the decay of the mean moisture in the system. In \sect{sec:forced}, we consider a cellular circulation in a bounded domain with a moisture source at the bottom boundary. This setup roughly resembles the Hadley cell \cite{Hadley1735}. We discuss the general features of the statistically steady humidity distribution and their dependence on the circulation strength. In the strong circulation limit, we derive an expression for the specific humidity PDF from which diagnostics such as evaporation rate and precipitation rate are obtained. \Sect{sec:conclude} concludes the paper.

\section{The advection--condensation model}
\label{sec:model}

Consider an ensemble of moist air parcels passively advected by a velocity field in a two-dimensional domain. When the specific humidity $Q$ of an air parcel at position $\V X=(X,Y)$ exceeds the local value of the saturation specific humidity $q_s(\V X)$, the excessive moisture condenses and precipitates out of the system. To a very good approximation, $q_s$ is proportional to the saturation vapour pressure which varies with temperature according to the Clausius--Clapeyron relation \cite{Andrews10}. Assuming that the temperature is independent of $x$ and decreases linearly with $y$,  $q_s$ decays exponentially in $y$ \cite{Pierrehumbert07}. Thus for the rest of this paper, we take
\begin{equation}
q_s(y) = \qmax\, e^{-\alpha y}.
\label{qs}
\end{equation}
for some constant $\alpha>0$.

Our goal is to investigate the effect of a large-scale circulation on the distribution of moisture. To this end, the prescribed velocity in our model is composed of a deterministic part $\V u=(u,v)$ representing large-scale coherent motions and a stochastic, $\delta$-correlated in time (white noise) component which mimics the small-scale random transport of the air parcels. Hence, the Lagrangian formulation of our advection--condensation model takes the form of a set of stochastic differential equations for the random variables $(X,Y,Q)$:
\begin{subequations}
\begin{align}
\dd X(t) &= u(X,Y)\, \dd t + \sqrt{2\kappa}\,\dd W_1(t), \label{xt} \\
\dd Y(t) &= v(X,Y)\, \dd t + \sqrt{2\kappa}\,\dd W_2(t), \label{yt} \\
\dd Q(t) &= [\mathcal S(X,Y)-\mathcal C(Y,Q)]\, \dd t.
\label{qt}
\end{align}
\label{sde}%
\end{subequations}
The Brownian motion of the parcel is modelled via the Wiener processes $W_1(t)$ and $W_2(t)$ with  diffusivity $\kappa$. $\mathcal S$ represents a moisture source. Generally, the condensation sink $\mathcal C$ is given by
\begin{equation}
\mathcal C = \tau_c^{-1}[Q-q_s(Y)]\,\step[Q-q_s(Y)]
 \,,\end{equation}
where $\tau_c$ is the condensation time scale and $\step$ denotes the Heaviside step function. Following previous studies \cite{Pierrehumbert07,OGorman06,Sukhatme11}, we take the rapid condensation limit $\tau_c\rightarrow 0$. Effectively, this means that $Q$ is reset to $q_s(Y)$ whenever the former exceeds the latter,
\begin{equation}
\mathcal C : Q(t) \mapsto \min\big\{Q(t),q_s[Y(t)]\big\}.
\label{cc}
\end{equation}
The specific form of $\mathcal S$ and $(u,v)$ will be given in the following sections when we consider initial-value  and  steady-state problems.

The two-dimensional model described above can represent an isentropic surface in the mid-troposphere with $x$ the distance in the east-west direction and $y$ the distance from the equator. The present setup can also be considered as a crude model for moisture transport by an overturning circulation in the free troposphere. Then $x$ represents the latitude or longitude and $y$ is the altitude. Generally, the typical length scales in the $x$- and $y$- directions are different. Here, it is understood that $X$ and $Y$ have been scaled by their respective typical length scales. For simplicity the re-scaled diffusivities in the two directions are assumed equal.

\section{Initial-value problem}
\label{sec:drying}

Let us consider a patch of initially saturated air in an unbounded domain with no moisture source, $\mathcal S=0$. Condensation may occur when individual air parcels move in the $y$-direction hence reducing the total moisture content in the system. We are interested in how a vortex, taken to be the solid-body rotation
\begin{equation}
u = -\Omega\, y \,,\quad v = \Omega\, x,
\label{vortex}
\end{equation}
with constant $\Omega$, added to the random motion of the air parcels modifies the drying process. Later, we shall see that the present setup is relevant to the emergence of a dry zone in the forced problem considered in \sect{sec:forced}. With typical length $\alpha^{-1}$ set by $q_s$ in \Eq{qs} and typical velocity $\Omega\alpha^{-1}$, the inverse P\'eclet number
\begin{equation}
\epsilon = {\kappa\alpha^2}/{\Omega}
\end{equation}
measures the importance of random motion relative to the circulation.

\begin{figure*}
\centering
\includegraphics[width=\textwidth]{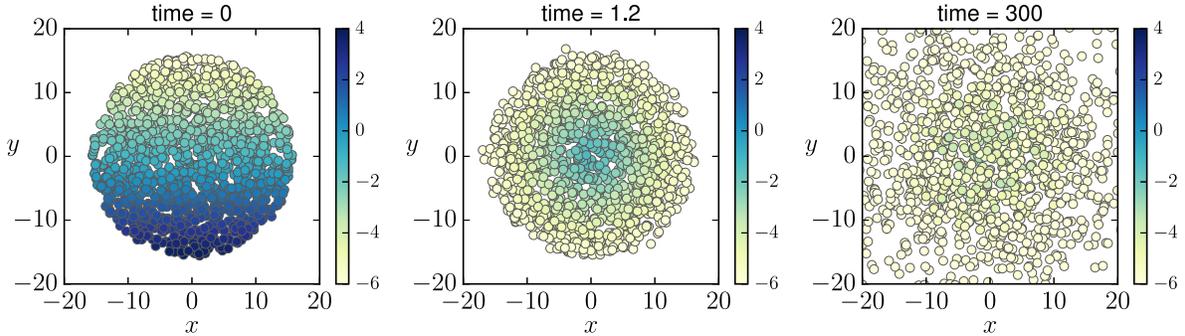}
\caption{Monte Carlo simulation of the stochastic advection-condensation model \Eq{sde} for the initial-value problem in \sect{sec:drying}. The position of air parcels is shown at different times with a colour scale that indicates the specific humidity $\log_{10}\!Q$ carried by each parcel. Values of the simulation parameters are given in the text.}
\label{drying}
\end{figure*}

\FF{drying} shows a typical Monte Carlo simulation of \Eq{sde} using the Euler--Maruyama method \cite{Higham01} for an ensemble of $10^7$ air parcels. The simulation parameters are $\Omega=5$, $\kappa=10^{-1}$, $\alpha=\ln(10)/\pi$ and $\qmax=0.1$. This gives $\epsilon\approx 0.01 \ll 1$, so this case is in the fast circulation limit. The value of $\alpha$ mimics the situation in the troposphere where the saturation specific humidity varies by several order of magnitudes with altitude as well as between the tropical and polar regions \cite{Sherwood10}. The parcels are initially distributed evenly over a circular area centred at the origin with radius $R=5\pi$ (left panel of \Fig{drying}). We are interested in a large patch $R \gg \alpha^{-1}$. Generally, we find that the drying process consists of a fast advective stage and a slow stochastic stage. We discuss these two stages in the following sections.

\subsection{Advective drying}
\label{advdry}

Initially, at $t=0$, all parcels are saturated and have $Q(0)=q_s[Y(0)]$. At $t=0^+$, the air parcels start to move in the counterclockwise sense along the circular streamlines of $(u,v)$ with small random fluctuations induced by the Brownian motion. The parcels that move in the $-y$ direction are entering regions where $q_s(Y) > Q$, thus no condensation occurs and $Q$ remains constant. On the other hand, for parcels moving in the $+y$ direction along a streamline of radius $r$, condensation starts immediately. These parcels continue to lose water vapour as condensation goes on until they reach $(X,Y) \approx (0,r)$ and $Q\approx q_s(r)$---the minimum of $q_s$ on the streamline. By the time
\begin{equation}
t_a \equiv 2\pi\Omega^{-1}
\label{ta}
\end{equation}
every parcel has made one complete revolution and a large amount of moisture has been lost: the moisture distribution becomes more or less axisymmetric with
\begin{equation}
Q(t_a) \approx q_s \big[\sqrt{X^2(0)+Y^2(0)}\big]
\end{equation}
for each parcel (middle panel of \Fig{drying}). The rapid initial drying is best exhibited by the decay of the {\it global specific humidity} defined as
\begin{equation}
\bar Q(t) \equiv \frac{1}{N}\sum_{i=1}^N Q_i(t),
\label{qbardef}
\end{equation}
where the sum is over all $N$ air parcels. Assuming all parcels have the same air mass, $\bar Q$ is simply the ratio of the total moisture mass to the total air mass in the system. During the advective drying stage, $\bar Q$ drops rapidly from its initial value at $t=0$ to
\begin{equation}
\bar Q(t_a) \approx \frac{1}{\pi R^2}\int_0^{2\pi}\!\!\!\dd\theta\int_0^{\infty}\!\!\qmax e^{-\alpha r}r\,\dd r = \frac{2\qmax}{(\alpha R)^2}
\label{qadv}
\end{equation}
at $t\approx t_a$. Further drying from this time on relies on the Brownian motion of the parcels and corresponds to the slow stochastic drying phase. \FF{qbar}(a) shows this transition for different $\Omega$ including the case without a vortex ($\Omega=0$), other simulation parameters are the same as in \Fig{drying}.
\begin{figure}
\centering
\includegraphics[width=0.9\textwidth]{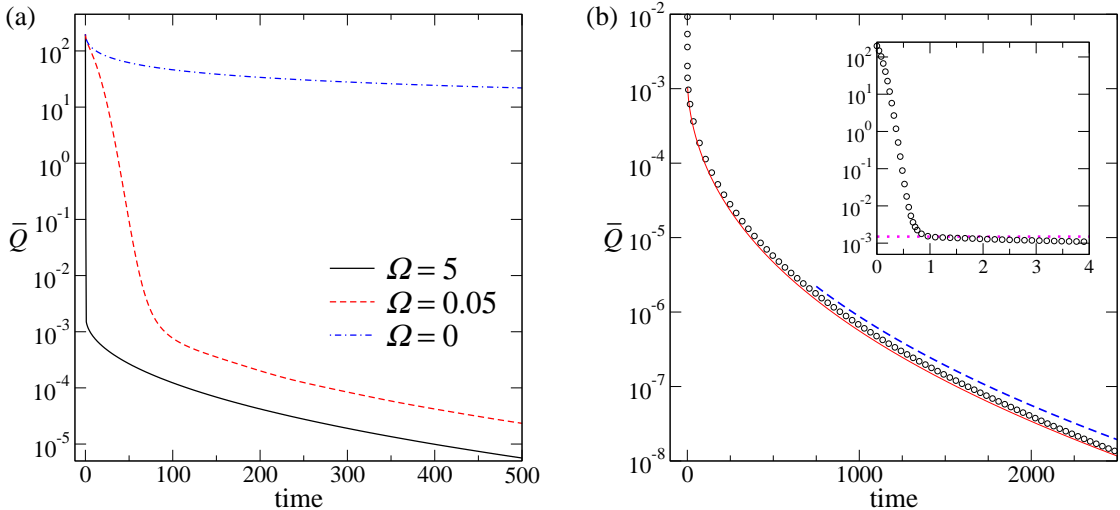}
\caption{Decay with time of the global specific humidity $\bar Q$ defined in \Eq{qbardef} for the initial-value problem in \sect{sec:drying}. (a) Transition from advective to stochastic drying for different values of angular velocity $\Omega$. (b) Monte Carlo simulation results (circles) for $\Omega=5 $ ($\epsilon\approx0.01$) are compared with the theoretical prediction in \Eq{qbar2} (solid line) and with the long-time asymptotic formula \Eq{asym} (dashed line). The inset shows the initial rapid decay, with the dotted line indicating the value at the end of  advective drying stage, $\bar Q(t_a)$  in \Eq{qadv}.}
\label{qbar}
\end{figure}

\subsection{Stochastic drying}
\label{stodry}

In the stochastic drying phase, an air parcel on a streamline of radius $r_1$ that wanders onto another streamline of radius $r_2>r_1$ is being quickly advected into the region of $y\approx r_2$ with lower saturation specific humidity. Rapid condensation within this region reduces the specific humidity of the parcel from $Q=q_s(r_1)$ to $Q=q_s(r_2)$. Our primary goal in this section is to calculate the resulting PDF of $Q$. Following previous work \cite{OGorman06,Pierrehumbert07}, this is achieved by considering the maximum excursion statistics of an air parcel.

Define the maximum excursion (in the $y$-direction) at time $t$ of an air parcel as
\begin{equation}
\Lambda(t) = \max_{s\in[0,t]}Y(s).
\end{equation}
Because of rapid condensation \Eq{cc}, the specific humidity of an air parcel at time $t$ is the minimum $q_s$ it encounters during the time interval $[0,t]$. Since $q_s$ decreases monotonically with $y$, this implies that the random variables $Q$ and $\Lambda$ are related by
\begin{equation}
Q(t) = q_s[\Lambda(t)] = \qmax e^{-\alpha\Lambda(t)}.
\label{ql}
\end{equation}
We first derive the equation satisfied by the cumulative distribution function $\cdf$ of $\Lambda$ for an air parcel located at $\V x$ at time $t$. Suppose there is an absorbing barrier at $y=\lambda$. We follow a parcel {\it backward} in time according to \Eqs{xt}{yt} and remove it from the system if its trajectory $\V X(t)$ hits the absorbing barrier at some $t>0$. It then follows that
\begin{equation}
\cdf \equiv \mathbb{P}[\Lambda(t) < \lambda | \V X(t)=\V x\,] = \mathbb{P}[\V X(0) \in \mathsf S | \V X (t)=\V x\,] = \mathbb E\{\chis[\V X(0)]\,|\V X (t)=\V x\,\},
\label{cdf}
\end{equation}
where $\mathsf S=(-\infty,\infty)\times(-\infty,\lambda]$, $\mathbb{P}(E_1|E_2)$ and $\mathbb E({E_1|E_2})$ denote respectively the probability and expectation of event $E_1$ conditioned upon event $E_2$, and $\chis$ is the indicator function
\begin{equation}
\chis(\V x) =
\begin{cases}
1 & \text{if $\V x \in \mathsf S$}, \\
0 & \text{if $\V x \not\in \mathsf S$}.
\end{cases}
\end{equation}
Since  backward trajectories are equivalent to forward trajectories under a reversal of $\V u$, \Eq{cdf} gives
\begin{equation}
\cdf = \mathbb E\{\chis [\V X(t)]\,|\V X (0)=\V x\,\}\big|_{\V u \mapsto -\V u}
\label{echis}
\end{equation}
and it follows that $\cdf$ satisfies the backward Kolmogorov equation \cite{Gardiner09,Pavliotis14},
\begin{equation}
\partialt{C} = -\V u \cdot \nabla C + \kappa \nabla^2 C,
\label{bke}
\end{equation}
with the boundary condition $\cdf=0$ at $y=\lambda$. The initial condition is $C(\lambda|\V x;0)=\step(\lambda-y)$ and we adopt the convention $\step(0)=0$ for the Heaviside function.

We now solve \Eq{bke} perturbatively for $C$ in the fast flow limit $\epsilon \ll 1$. Nondimensionalising using $\V x=\alpha^{-1}\hat{\V x}$, $t=(\alpha^{-2}\kappa^{-1})\hat t$, and $\V u=(\alpha^{-1}\Omega)\hat{\V u}$, then suppressing the hats, \Eq{bke} becomes
\begin{equation}
\partialt{C} = -\epsilon^{-1} \V u \cdot \nabla C +  \nabla^2 C.
\label{bke1}
\end{equation}
Adopting  polar coordinates, we expand
\begin{equation}
C(\lambda|r,\theta;t) = C_0 + \epsilon^{1/2} C_1 + \epsilon C_2 + \cdots,
\end{equation}
where the powers of $\epsilon^{1/2}$ turn out to be required for matching with a boundary layer around $r=\lambda$. At the leading order $\epsilon^{-1}$, we find that
\begin{equation}
\V u \cdot \nabla C_0 = 0,
\end{equation}
which means $C_0=C_0(\lambda|r;t)$ is constant along  streamlines. Hence, at the lowest order, the moisture distribution is axisymmetrized by $\V u$, as described in \sect{advdry}. The next-order solution is similarly axisymmetric $C_1=C_1(\lambda|r;t)$. At $O(\epsilon^0)$, we obtain 
\begin{equation}
\partialt{C_0} = -\V u \cdot \nabla C_2 +  \nabla^2 C_0.
\label{ep1}
\end{equation}
For the solid-body rotation \Eq{vortex}, $\V u \cdot \nabla C_2=\Omega\partial_\theta C_2$. Hence averaging \Eq{ep1} over $\theta$ eliminates the term involving $C_2$, leading to the one-dimensional heat equation for $C_0$ (in dimensional variables)
\begin{equation}
\partialt{C_0} = \frac{\kappa}{r}\partialr{}\left(r\partialr{C_0}\right).
\label{f0}
\end{equation}
For fast circulation, the boundary and initial conditions of $C$ implies $C_0(\lambda|r;t)=0$ for $r\geqslant\lambda$ and $C_0(\lambda|r;0)=\step(\lambda-r)$. The solution $C_0$ obtained in this manner has discontinuous derivatives at $t=0$ and $r=\lambda$. These are smoothed out in boundary layers: a boundary layer in time of size $O(\epsilon)$ matches with the advective drying solution described in \sect{advdry}; a boundary layer around $r=\lambda$ of size $O(\epsilon^{1/2})$ where radial diffusion is important ensures a smooth transition between the positive values of $C$ for $r < \lambda$  and zero values for $r \geqslant \lambda$. The details of the solution within the boundary layers are unimportant for $C_0$ outside and we do not consider them further. Solving \Eq{f0} for $C_0$, we obtain the PDF of the maximum excursion for a parcel landing at $r$ at time $t$,
\begin{align}
P_\Lambda(\lambda|r,t) = \frac{\partial C_0}{\partial\lambda}
= \frac{2r}{\lambda^2}\! \sum_{n=1}^{\infty}\!\frac{1}{J_1(z_n)}
\!\!\left[
\frac{2z_n \kappa t}{\lambda r}J_0\!\left(\frac{z_n r}{\lambda}\right)
+ J_1\!\left(\frac{z_n r}{\lambda}\right)
\right]\!
\exp\!\left(-\frac{z_n^2 \kappa t}{\lambda^2}\right),
\label{pdfb}
\end{align}
for $r < \lambda$. Here, $J_0$ and $J_1$ are the zeroth and first order Bessel functions of the first kind, and $z_n$ is the $n$th zero of $J_0$. Using \Eq{ql}, we finally have the leading-order PDF of the specific humidity for a parcel arriving at position $r$ at time $t$,
\begin{align}
P_Q(q|r,t) = \frac{-2\hat r}{\qmax\,\hat q\ln^2\!\hat q}
\sum_{n=1}^{\infty}\frac{1}{J_1(z_n)}
\left[
\frac{2z_n\hat t}{\hat r\ln\hat q}J_0\!\left(\frac{z_n\hat r}{\ln\hat q}\right)
+ J_1\!\left(\frac{z_n\hat r}{\ln\hat q}\right)
\right]
\exp\!\left(-\frac{z_n^2\,\hat t}{\ln^2\!\hat q}\right),
\label{pdfq}
\end{align}
for $\hat q<e^{-\alpha r}$ where $\hat q=q/\qmax$, $\hat r=\alpha r$ and $\hat t=\alpha^2\kappa t$.

\begin{figure}
\centering
\includegraphics[width=0.96\textwidth]{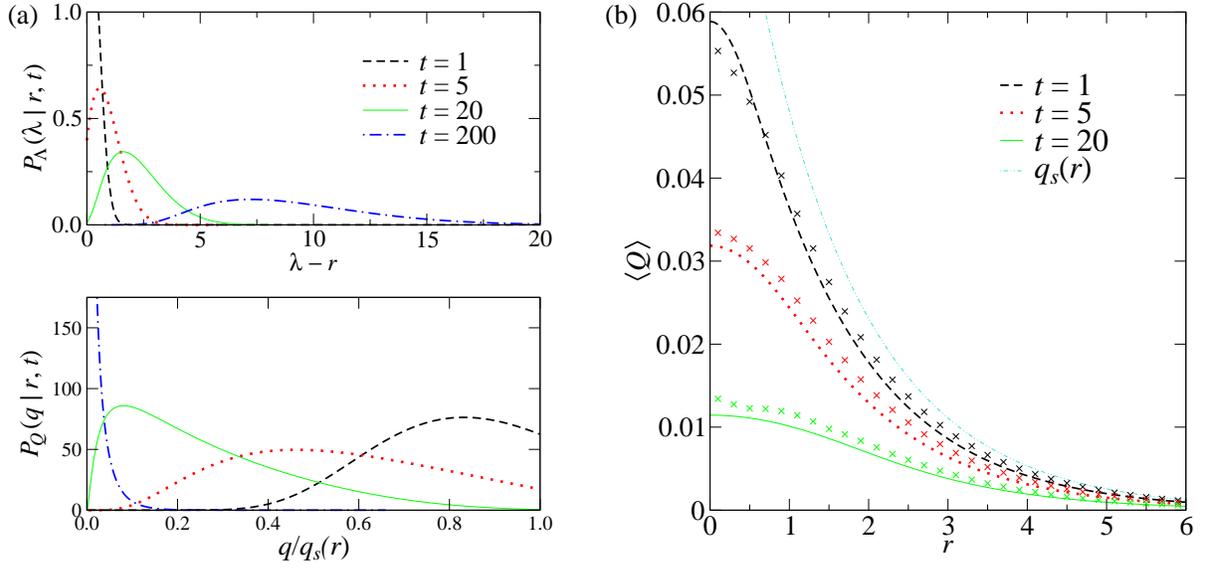}
\caption{Moisture distribution for the initial-value problem: (a) Theoretical PDFs for the maximum excursion $\Lambda$ and the specific humidity $Q$ of a parcel at $r=\pi/2$ and different times $t$. To ensure convergence in the tails, these PDFs are calculated from \Eq{pdfb} and \Eq{pdfq} by including the first 50 terms of the series. Same parameter values as those in the Monte Carlo simulation of \Fig{drying} are used in the formulas. (b) Profile of ensemble mean specific humidity $\avg{Q}$ at different times $t$. The curves are theoretical prediction calculated using \Eq{qavgdef} and the first 10 terms in  expression \Eq{pdfq} for $P_Q(q|r,t)$. The crosses are results from Monte Carlo simulations. The saturation specific humidity profile $q_s(r)$ is also shown.}
\label{pdfqb}
\end{figure}

Using parameters matching those of \Fig{drying}, \Fig{pdfqb}(a) plots \Eq{pdfb} and \Eq{pdfq} for $r=\pi/2$ at different times $t$. At early times, most air parcels have not moved far  from their initial position. So a parcel landing at $r$ is most likely coming from the vicinity of $r$, implying its maximum excursion is either equal to or only slightly larger than $r$, hence its specific humidity is equal to or slightly less than $q_s(r)$. As time goes by, more and more parcels have visited places with small $q_s$ and undergone condensation before arriving at $r$. Thus, the peak of $P_\Lambda(\lambda|r,t)$ shifts to larger $\lambda$ while that of $P_Q(q|r,t)$ shifts to  smaller $q$.

We now compare predictions of our theory to results from the Monte Carlo simulation described in \Fig{drying} which has $\epsilon\approx0.01$. The fast circulation limit ($\epsilon\rightarrow 0$) assumed in the theory means that the moisture field is axisymmetrized instantaneously at $t=0^+$. However, it always takes a finite amount of time, namely $t_a = 2 \pi \epsilon$ [see the dimensional \Eq{ta}], for that to happen in a simulation with small but finite $\epsilon$. We will therefore compare theoretical prediction at time $t$ to the corresponding numerical results at time $t+t_a$. 

We first look at the spatial profile of the mean specific humidity 
\begin{equation}
\avg{Q}(r,t) = \int_0^{\qmax e^{-\alpha r}}\!\!\! q\, P_Q(q|r,t)\, \dd q.
\label{qavgdef}
\end{equation}
\FF{pdfqb}(b) compares simulation results for $\avg{Q}$ at different times with the theoretical prediction calculated from \Eq{pdfq}. The numerical estimate of $\avg{Q}$ at a given $r$ is obtained by averaging the specific humidity $Q$ over all the parcels located within a thin annulus of radii $r\pm\delta$ with $\delta=0.05$. Note that our theoretical prediction assumes the parcels are initially distributed uniformly across the $(x,y)$-plane while the Monte Carlo simulation initialises parcels inside a circle of radius $R$ only. However since $R \gg \alpha^{-1}$ and $\epsilon \ll 1$, the parcels that are not sampled make a negligible contribution to the statistics. We find reasonable agreement between the theoretical and numerical results with the largest discrepancy near $r=0$. This is due to the lack of data points and the deviation from the fast circulation limit near $r=0$ (recall $|\V u|=\Omega r$).

We can also predict the decay of the global specific humidity $\bar Q(t)$, defined in \Eq{qbardef}, for a patch of initially saturated parcels. The circles in \Fig{qbar}(b) shows $\bar Q(t)$ measured in the simulation. For $N$ parcels distributed uniformly in the $(x,y)$-plane with number density $\rho=N/(\pi R^2)$, the expectation value of $\bar Q$ can be calculated as
\begin{equation}
\avg{\bar Q}(t) = \frac{1}{N} \int_0^{\qmax}\! q \, \dd q \iint P_Q(q|r,t) \rho \, \dd x \dd y
= \frac{1}{\pi R^2} \int_0^{\qmax}\!  q \, \dd q \int_0^{2\pi}\!\!\!\dd\theta \int_0^{-\frac{1}{\alpha}\!\ln\hat q} r P_Q(q|r,t)\, \dd r.
\label{qbar1}
\end{equation}
Performing the spatial integration, we obtain
\begin{equation}
\avg{\bar Q}(t) = \frac{-4\qmax}{\alpha^2 R^2}\sum_{n=1}^\infty \int_0^1 \exp\!\left(-\frac{z_n^2\,\hat t}{\ln^2\!\hat q}\right)
\left[\frac{2\hat t}{\ln\hat q} + \frac{J_2(z_n)}{z_nJ_1(z_n)}\ln\hat q \right] \dd \hat q.
\label{qbar2}
\end{equation}
In contrast to $P_\Lambda$, $P_Q$ or $\avg{Q}$ in \Fig{pdfqb}, \Eq{qbar2} for $\avg{\bar Q}$ is dominated by the first term which we plot as a solid line in \Fig{qbar}(b). We see that the theory is in good agreement with the Monte Carlo simulation. The long-time decay of $\avg{\bar Q}$ can be found from \Eq{qbar2} using  Laplace's method as detailed in \ref{larget}. The result, also plotted in  \Fig{qbar}(b), is
\begin{equation}
\avg{\bar Q}(t) \propto t^{5/6}\exp\left[-\frac{3}{2^{2/3}}\big(z_1^2\alpha^2\kappa t\big)^{1/3}\right]
\text{ as } t\rightarrow\infty.
\label{asym}
\end{equation}

\subsection{A general incompressible flow}

In this section, we outline an extension of the above calculation to arbitrary flows with closed streamlines. A motivation for this extension is that the transport of moisture in mid-latitudes is primarily along moist isentropic surfaces. Such transport is driven by large-scale baroclinic eddies  and can roughly be modelled by a wavy velocity field in a periodic channel which our extension covers.

The main idea is to generalise the polar coordinates $(r,\theta)$ used for axisymmetric flows to the pair $(\psi,\tau)$ where $\psi$ is the value of the streamfunction and $\tau$ is the elapsed time along a streamline defined by
\begin{equation}
\tau = \int_\psi \frac{\dd l}{|\nabla \psi|},
\end{equation} 
where $l$ is the arclength and the integral is along a streamline. The advective phase of the drying reduces the humidity of air parcels initially located on a streamline $\psi$ to $Q=q_s(y_\psi)$, where $y_\psi$ denotes the maximum value of $y$ along the streamline. To analyse the later phase of stochastic drying, we need to consider the backward Kolmogorov equation \Eq{bke1} for $C(\lambda|\psi,\tau;t)$. To leading order this reduces to
\begin{equation}
\V u\cdot\nabla C_0 =  \frac{\partial C_0}{\partial \tau} = 0,
\end{equation}
which implies that $C_0=C_0(\lambda|\psi;t)$. Introducing this into \Eq{ep1} and averaging over $\tau$ yields
\begin{equation}
\partialt{C_0} = \kappa\frac{\partial}{\partial\psi}
\left(\oint_\psi|\nabla\psi|\,\dd l\, \frac{\partial C_0}{\partial\psi}
\right)
\label{genflow}
\end{equation}
(see \cite{Rhines83} and Appendix A of \cite{Haynes14} for details). This heat-like equation, which reduces to \Eq{f0} for a solid body rotation, can be solved (numerically in general) with the initial condition $C_0(\lambda|\psi;0)=1$ if $y_\psi < \lambda$ and $0$ otherwise. The PDFs $P_\Lambda(\lambda |\psi,t)$ and $P_Q(q|\psi,t)$ follow.

\section{Steady-state problem}
\label{sec:forced}

Water vapour condensed and precipitated out of the atmosphere is replenished by evaporation of liquid water from the oceans and land. As mentioned in \sect{intro}, the large-scale cycling of atmospheric water can sometimes be viewed as taking place inside a single overturning cell \cite{Pauluis10,Sherwood10}. One question that naturally arises is: how does the moisture distribution within the cell change with the strength and other properties of the circulation? Here, we investigate this within the context of the advection--condensation paradigm.

\begin{figure}
\centering
\includegraphics[width=\textwidth]{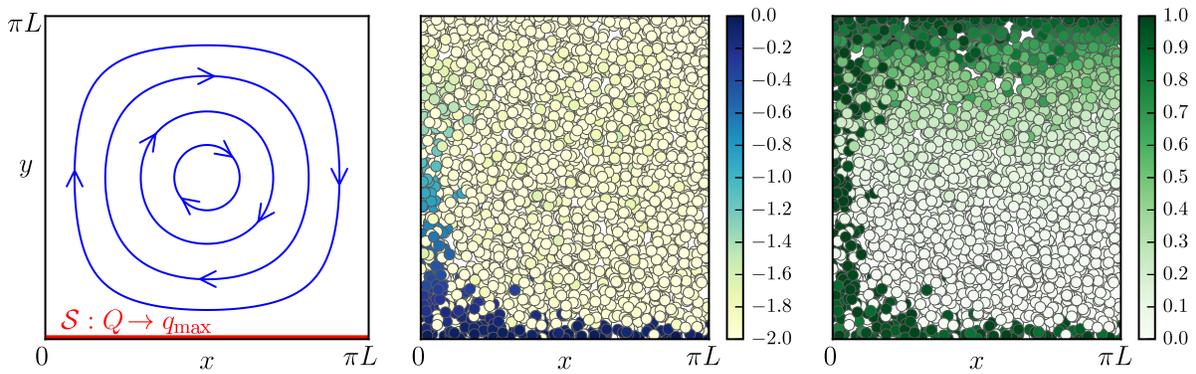}
\caption{Steady-state problem: (left) schematics showing the streamlines \Eq{stream} and the moisture source at the bottom boundary; (middle and right) snapshots from a Monte-Carlo simulation of \Eq{sde} with $\epsilon=10^{-2}$ showing the positions of the moist parcels, the colour scale indicates the value of $\log_{10}\!Q$ (middle) and relative humidity $Q/q_s(Y)$ (right).}
\label{cell_qs}
\end{figure}

As a simple representation of an overturning cell, we consider the velocity $(u,v)=(-\partial_y\psi,\partial_x\psi)$ given by the streamfunction
\begin{equation}
\psi(x,y) = U L \sin(x/L) \sin(y/L)
\label{stream}
\end{equation}
in a bounded domain $[0,\pi L] \times [0,\pi L]$ with reflective boundary condition, see \Fig{cell_qs}. Recall that $x$ and $y$ are re-scaled to have the same typical length $L$ as discussed near the end of \sect{sec:model}. The evaporation source is modelled as a boundary condition at $y=0$: the specific humidity $Q$ of air parcels hitting (and reflecting on) the bottom boundary is reset to $\qmax$, the saturation value there \cite{Pierrehumbert07,Sukhatme11}. The saturation profile is given by \Eq{qs}. From here on, we fix $\qmax=q_s(0)=1$ and $\qmin\equiv q_s(\pi L)=0.01$. The fate of a moist parcel under the action of large-scale circulation \Eq{stream}, Brownian motion and condensation is then governed by \Eq{sde}. If we interpret $x$ as the meridional direction and $y$ as altitude, this setup resembles the Hadley cell.

\begin{figure}
\centering
\includegraphics[width=\textwidth]{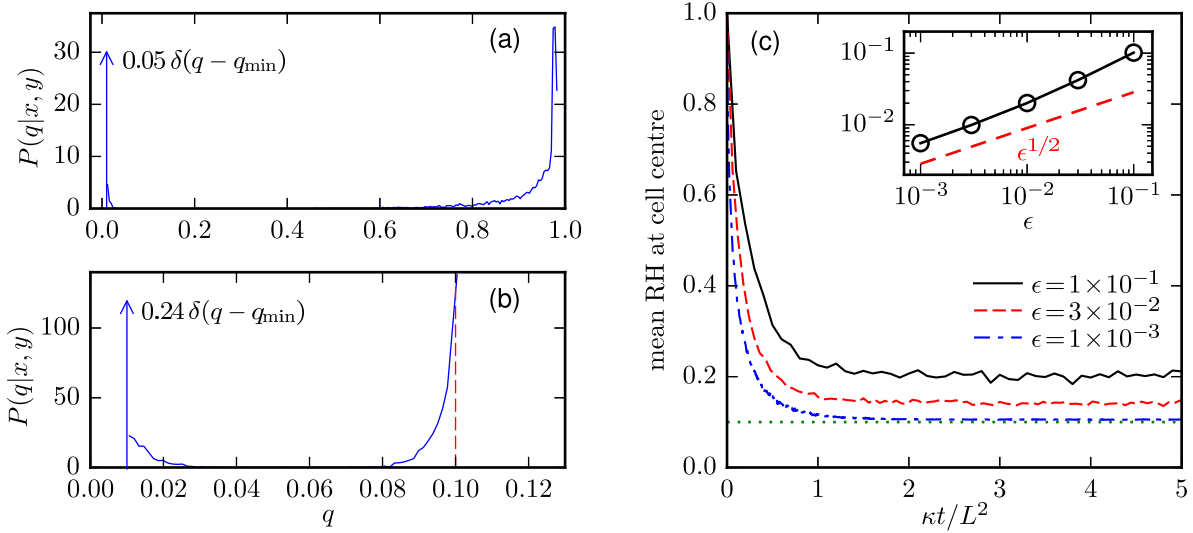}
\caption{(a) Conditional PDF $P(q|x,y)$ of specific humidity inside the source boundary layer at $(x,y)=(L/2,L/200)$ for $\epsilon=3\times10^{-2}$. The PDF consists of a continuous part (solid line) and a dry spike indicated by an arrow at $q=\qmin$. (b) As in (a) but for $(x,y)=(L/200,L/2)$ inside the condensation boundary layer. The dashed line marks the value of $q_s(L/2)$. (c) Mean relative humidity (RH) for parcels in the neighbourhood (within a distance of $L/80$) of cell centre as a function of time $t$. For small $\epsilon$, its long-time value RH$^{\rm ctr}_\infty$ approaches the minimum attainable value RH$^{\rm ctr}_{\min}=\qmin/q_s(L/2)=0.1$ (dotted line). The inset shows the difference RH$^{\rm ctr}_\infty-$ RH$^{\rm ctr}_{\min}$ scaling as $\epsilon^{1/2}$.}
\label{qcpdfbl}
\end{figure}

\FF{cell_qs} shows a snapshot of the statistically steady state attained in a Monte-Carlo simulation of the system. The domain is initially saturated. For all simulations presented in this section, we use $10^6$ parcels. We focus on situations when the circulation is strong, with the inverse P\'eclet number
\begin{equation}
\epsilon = \kappa/(UL) \lesssim 1.
\end{equation}
Our simulations show that there are generally three distinct regions inside the cell:
\begin{enumerate}
\item {\it The source boundary layer}. Since the circulation is tangential to the boundary, it is by means of the small-scale Brownian motion that the parcels hit the bottom boundary and moisture is injected into the domain. When the vertical random motion of the recently-saturated parcels near the source is balanced by the sweeping (toward $x=0$) of the circulation, a boundary layer of high humidity is formed at $y=0$. Interestingly, as can be seen in \Fig{cell_qs}, mixed inside this layer of mostly wet parcels are parcels with $Q\approx\qmin$ that subsides from aloft. This results in a bimodal local PDF $P(q|x,y)$ for the specific humidity inside this boundary layer, see \Fig{qcpdfbl}(a).

\item {\it The condensation boundary layer}. Advected by the circulation, the wet parcels with $Q\approx\qmax$ in the source boundary layer converge toward a narrow region near $x=0$ before moving upward. The water vapour in these parcels then quickly condenses as $q_s$ decreases, keeping the relative humidity $Q/q_s(Y)\approx 1$ (\Fig{cell_qs}). Such a region of intense precipitation is reminiscent of the Intertropical Convergence Zone. \FF{qcpdfbl}(b) shows a typical PDF of $Q$ inside this boundary layer. Similar to (i), the dry parcels brought in by the Brownian motion give $P_Q(q|x,y)$ two peaks at $q=\qmin$ and $q=q_s(y)$.

\item {\it The dry interior}. The bulk interior (as well as the top boundary and the descending arm) of the cell is mainly occupied by parcels with $Q\approx\qmin$, creating a patch of relative humidity minimum \cite{OGorman11} about the cell centre. This is because parcels that pick up moisture from the source are quickly advected by the circulation around the periphery, leaving the interior largely oblivious of the source. The upshot is the inner region losing its moisture through advective and stochastic drying (\sect{sec:drying}). \FF{qcpdfbl}(c) shows the decrease of relative humidity at the centre of the cell with time. The equilibrium mean specific humidity inside the dry patch is maintained slightly above $\qmin$ by moisture mixing in \cite{Pierrehumbert98a} from the condensation and source boundary layers via Brownian motion.

\end{enumerate}
Through the interplay between coherent stirring and small-scale random motion, our idealised model develops the interesting features of boundary layer and relative humidity minimum. This is in contrast to a one-dimensional system of Brownian parcels \cite{Sukhatme11}. With the qualitative picture described above in mind, we examine quantitatively how the strength of the circulation controls the system in the next sections.

\subsection{Water vapour PDF in the fast circulation limit}
\label{sec:pdf}

We first derive the steady joint PDF $P(q,x,y)$ governing the equilibrium statistics of the parcel position and specific humidity in \Eq{sde}. The steady Fokker--Planck equation satisfied by $P(q,x,y)$ is
\begin{equation}
\vec u \cdot \nabla P + \partial_q \left[ (\mathcal S-\mathcal C) P \right] = \kappa\nabla^2 P.
\label{sFP}
\end{equation}
Following Sukhatme \& Young \cite{Sukhatme11}, rapid condensation \Eq{cc} implies that we only need to consider \Eq{sFP} in a region of the $(q,x,y)$-space where $\qmin\leqslant q\leqslant q_s(y)$. Within this region, $\mathcal S=\mathcal C=0$. We are interested in the limit of fast circulation. Thus, upon re-scaling $\vec x=L \hat{\vec x}$, $\vec u = U\hat{\vec u}$ and suppressing the hats, we consider
\begin{equation}
\vec u \cdot \nabla P = \epsilon\nabla^2 P
\label{eFP}
\end{equation}
with $\epsilon \ll 1$. From \Eq{eFP}, it follows \cite{Sukhatme11} that the marginal PDF
\begin{equation}
p(x,y) = \int_{\qmin}^{q_s(y)} P(q,x,y)\, \dd q = \frac{1}{\pi^2}.
\label{marpdf}
\end{equation}
This simply means the number density of parcels is uniform over the entire cell. The boundary conditions are no-flux at all edges except the bottom one at which $P(q,x,0) = \pi^{-2} \delta(q-\qmax)$ representing the source. We also know that $P(q,x,\pi)=\pi^{-2} \delta(q-\qmin)$ because when a parcel hits $y=\pi$ (and then subsides), it has a probability one that $Q=\qmin$. This idealisation at the top edge coupled with a localised boundary source implies that $P(q,x,y)$ generally contains a singular dry spike \cite{Sukhatme11} at $q=\qmin$, as exemplified in \Fig{qcpdfbl}, in addition to a continuous part $F(q,x,y)$:
\begin{equation}
P(q,x,y) = \beta(x,y)\pi^{-2} \delta(q-\qmin) + F(q,x,y).
\label{spike}
\end{equation}

Equations of the form \Eq{eFP} have been widely studied in different areas such as magnetohydrodynamics \cite{Childress79}, transport in convective rolls \cite{Shraiman87,Young89} and two-dimensional vortex condensate \cite{Gallet13}. As $\epsilon\rightarrow 0$, it is well known that boundary layers of thickness $\eprt$ form around the periphery. Following standard procedures, we introduce the von Mises transformation \cite{vonMises27} $(x,y)\rightarrow(\psi,\gamma)$ where
\begin{equation}
\gamma \equiv \int_0^{l}|\nabla\psi(l')|\,\dd l'
\end{equation}
is the integral of the speed along the cell boundary. The speed is parametrized by the arclength $l$ and we choose $l=0$ at $(\pi,0)$ so that $\gamma=0$ at $(\pi,0)$, 2 at $(0,0)$, 4 at $(0,\pi)$ and 6 at $(\pi,\pi)$. The variables $\psi$ and $\gamma$ track the variation of $P$ across and along streamlines respectively. Inside the boundary layers, $\partial_\psi\sim\epsilon^{-1/2}$ and $\partial_\gamma\sim O(1)$ as advection along streamlines balances diffusion (of probability) across streamlines. We let $\psi=\eprt\hat\psi$ and substitute
\begin{equation}
P = P_0 + \eprt P_1 + \epsilon P_2 + \dots
\end{equation}
into \Eq{eFP}, we have to leading order in $\epsilon$,
\begin{align}
&\quad& &\qquad& \partialgma{P_0} = \partialpp{P_0} \qquad\text{(inside the boundary layers).}
\label{blFP}
\end{align}
On the other hand, in the cell interior outside the boundary layers as well as the corner regions, the Laplacian term in \Eq{eFP} is negligible to leading order \cite{Childress79,Gallet13}. Thus, we have
\begin{align}
&\qquad& &\qquad& \vec u \cdot \nabla P_0 = 0 \qquad\text{(for cell interior and corners).}
\label{inFP}
\end{align}
In the following, we derive $P_0$ for different regions of the cell.

\subsubsection{Cell interior, boundary layers at $x=\pi$ and $y=\pi$}

The amplitude of the dry spike in \Eq{spike} drops sharply from $\beta=1$ at $y=\pi$ to $\beta\approx1/2$ over a distance of $y\sim O(\eprt)$. Inside this top boundary layer, let $\hat Y=\epsilon^{-1/2}(\pi-y)\sim O(1)$. We then see that rapid condensation restricts the specific humidity to lie within
\begin{equation}
\qmin \leqslant Q \lesssim \qmax e^{-\alpha(\pi-\eprt\hat Y)} \approx \qmin(1+\alpha\eprt\hat Y).
\label{qminplus}
\end{equation}
Since $Q=\qmin$ to leading order for all parcels, we choose not to resolve the separation between the dry spike and the smooth contribution $F(q,x,y)$ to $P(q,x,y)$ and we take
\begin{equation}
P_0 = \pi^{-2}\delta(q-\qmin).
\label{ptop}
\end{equation}

We now turn to the cell interior. Eq.\ \Eq{inFP} implies that  $P_0=P_0(q,\psi)$ is constant along streamlines. To derive $P_0$, we consider \Eq{eFP} at order $\epsilon$: $\vec u\cdot\nabla P_2=\nabla^2 P_0$. Integrating this equation along streamlines [the same calculation that leads to \Eq{genflow}] gives the solvability condition for $P_2$:
\begin{equation}
\frac{\partial}{\partial\psi}
\left[ \Gamma(\psi) \frac{\partial P_0}{\partial\psi} \right] = 0
\quad\text{where}\quad
\Gamma(\psi) \equiv \oint_\psi|\nabla\psi|\,\dd l.
\end{equation}
The circulation $\Gamma$ increases monotonically from $\Gamma(0)=-8$ at the boundary to $\Gamma(1)=0$ at the centre \cite{Haynes14}. Hence, we conclude that for finite $\partial_\psi P_0$, $P_0$ must be independent of $\psi$. By matching to the boundary layer as $y\rightarrow\pi$, we see that $P_0$ in the cell interior is also given by \Eq{ptop}.

Inside the boundary layer near $x=\pi$ (where $\gamma=7+\cos y$), \Eq{ptop} provides the ``initial'' condition (at $\gamma=6$) for \Eq{blFP} because \Eq{inFP} in the corner regions ensures $P_0$ joins smoothly across neighbouring boundary layers. With zero-flux at the boundary $\hat\psi=0$ and matching to the interior solution \Eq{ptop} as $\hat\psi\rightarrow\infty$, it follows $P_0$ is once again given by \Eq{ptop}.

\subsubsection{Source boundary layer}
\label{aa}
We have seen in \Fig{qcpdfbl}(a) an example of the bimodality of extreme high and low specific humidity in the source boundary layer near $y=0$ (where $\gamma=1+\cos x$). From the discussion in the previous section, we know that the dry parcels flowing in from upstream and from the interior have $\qmin\leqslant Q<\qmin+O(\eprt)$. Following similar arguments, rapid condensation dictates that the specific humidity of the wet parcels lie between $\qmax$ and $\qmax-O(\eprt)$, we therefore write
\begin{equation}
P_0 = G(\hat\psi,\gamma) \pi^{-2}\delta(q-\qmin) + [1-G(\hat\psi,\gamma) ]\pi^{-2}\delta(q-\qmax) \quad\text{for}\quad 0 < \gamma < 2,
\label{bl1}
\end{equation}
making sure that \Eq{marpdf} is satisfied. We emphasise the distinction  
between $\beta$ in \Eq{spike} and $G$ in \Eq{bl1}: while $\beta$ describes parcels with $Q=\qmin$ exactly, $G$ is a leading-order approximation encompassing the range of $Q$ in \Eq{qminplus}.
From \Eq{blFP}, $G$ satisfies the heat equation
\begin{equation}
\partialgma{G} = \partialpp{G}.
\label{heat}
\end{equation}
The initial condition at $\gamma=0$ is obtained by joining the boundary layer upstream via the corner region at $x=\pi$. The source at the bottom edge $\hat\psi=0$ and matching to the interior as $\hat\psi\rightarrow\infty$ give the boundary conditions. Thus,
\begin{equation}
G(\hat\psi,0)=1, \quad G(0,\gamma)=0, \quad G(\infty,\gamma)=1
\end{equation}
and the solution is
\begin{equation}
G = \erf \left(\frac{\hat\psi}{2\sqrt{\gamma}}\right) 
= \erf \left(\frac{y\sqrt{1-\cos x}}{2\eprt}\right).
\label{gg1}
\end{equation}

\subsubsection{Condensation boundary layer}

The condensation layer near $x=0$ (where $\gamma = 3 - \cos y$) is the region of concentrated precipitation in the model. \FF{qcpdfbl}(b) shows a typical bimodal distribution of $Q$ in this layer. Dry parcels in \Eq{qminplus} once again contribute to the peak near $\qmin$. The peak at $q_s(y)$ has an $O(\epsilon)$ width extended toward $q<q_s(y)$ because some parcels at $y+y_1 (y_1>0)$ are able to random walk downward against $\vec u$ to reach $y$. We estimate the maximum $y_1$ by balancing upward advection and downward Brownian motion: $y_1\sim U\tau_1\sim\sqrt{\kappa\tau_1}$ (in dimensional variables) for some time $\tau_1$. This leads to $y_1\sim\kappa U\sim\epsilon L$ which implies these parcels have $Q\approx q_s(y)-O(\epsilon)$ due to rapid condensation. (This is consistent with the leading-order equation \Eq{blFP} which neglects along-flow diffusion.) The solution can therefore be written as
\begin{equation}
P_0 = G(\hat\psi,\gamma) \pi^{-2}\delta(q-\qmin) + [1-G(\hat\psi,\gamma) ]\pi^{-2}\delta[q-q_s(y_\gamma)] \quad\text{for}\quad 2 < \gamma < 4,
\label{bl2}
\end{equation}
where $y_\gamma = \cos^{-1}(3-\gamma)$. The function $G$ in the range $2 < \gamma < 4$  satisfies the heat equation \Eq{heat} with initial and boundary conditions
\begin{equation}
G(\hat\psi,2^+) = G(\hat\psi,2^-), \quad \partial_{\hat\psi} G(0,\gamma)=0, \quad 
G(\infty,\gamma)=1
\end{equation}
with $G(\hat\psi,2^-)$ obtained from \Eq{gg1}. The solution is given by
\begin{equation}
G = \erf\left(\frac{\hat\psi}{2\sqrt{\gamma}}\right)
+\frac{1}{\sqrt{\pi(\gamma-2)}}\int_0^\infty e^{-\frac{(\hat\psi'+\hat\psi)^2}{4(\gamma-2)}}
G(\hat\psi',2^-)\,\dd\hat\psi'.
\label{gg2}
\end{equation}
Note that the integral term above tends to zero as $\gamma\rightarrow2$ or $\hat\psi\rightarrow\infty$ as expected.

\subsection{Surface evaporation, boundary-layer ventilation and vertical flux}

Equipped with the joint PDF $P_0(q,x,y)$, we now study the transport of moisture from the source to the upper part of the domain. The mean specific humidity at position $(x,y)$ is given by the conditional expectation 
\begin{equation}
\avg{Q}(x,y) = \int_{\qmin}^{\qmax} q \tilde P(q|x,y)\, \dd q,
\end{equation}
with the conditional probability density
$\tilde P(q|x,y) = P(q,x,y)/p(x,y) = \pi^2 P(q,x,y)$.
The steady Fokker-Planck equation \Eq{sFP} implies the balance
\begin{equation}
\nabla\cdot\big(\avg Q\vec u - \kappa \nabla \avg Q\big)
= -\int_{\qmin}^{\qmax}\mathcal C\tilde P(q|x,y)\,\dd q \equiv -\avg{\mathcal C}
\label{fluxbal}
\end{equation}
for $y>0$. Apart from a factor of constant air density, $\avg{\mathcal C}$ is the mean moisture mass condensed per unit time per unit area. By integrating \Eq{fluxbal} over the region above a given $y$ and applying the divergence theorem, we find that the net upward transport of moisture mass across height $y$ per unit time is proportional to
\begin{equation}
\Phi(y) = \int_0^{\pi L} \left(v\avg Q - \kappa\partialy{\avg Q}\right) \dd x.
\label{phiy}
\end{equation}
We refer to $\Phi(y)$ as the vertical moisture flux.

The surface evaporation rate, i.e. the rate at which moisture is introduced by the source at $y=0$, is given by $\Phi(0)$. The idealisation of Brownian small-scale motion leads to air parcels continuously picking up and losing moisture by bouncing on and off the bottom edge multiple times in quick succession. This results in an infinite $\Phi(0)$, although much of this moisture is quickly lost in the immediate vicinity of $y=0$ \cite{Sukhatme11}. Thus for the present model, we focus on a more relevant measure of moisture input. We define the net surface evaporation rate $\Enet$ to be the surface moisture flux attributed only to the dry air parcels, specifically parcels with $Q<q_s(\eprt L)$. Because the approximation $P_0$ in \Eq{bl1} incorporates all parcels within an $\eprt$-neighbourhood of $\qmax$ into the spike at $\qmax$ which does not contribute to $\Phi(0)$, we can predict $\Enet$ by substituting \Eq{bl1} into \Eq{phiy} and evaluating the integral at $y=0$. Noting that $v$ vanishes on the bottom boundary, we have
\begin{equation}
\Enet = \sqrt{\kappa UL}\sqrt{\frac{8}{\pi}}\,(\qmax-\qmin)
\label{enet_th}
\end{equation}
and the dimensionless  $\Enet/(UL)\propto \eprt$. Hence the moisture input increases with the square root of the circulation strength. \FF{evap_flux}(a) shows good agreement between the theory and $\Enet$ obtained from a number of Monte-Carlo simulations over the range $10^{-3} \leqslant \epsilon \leqslant 10^{-1}$.

\begin{figure}
\centering
\includegraphics[width=\textwidth]{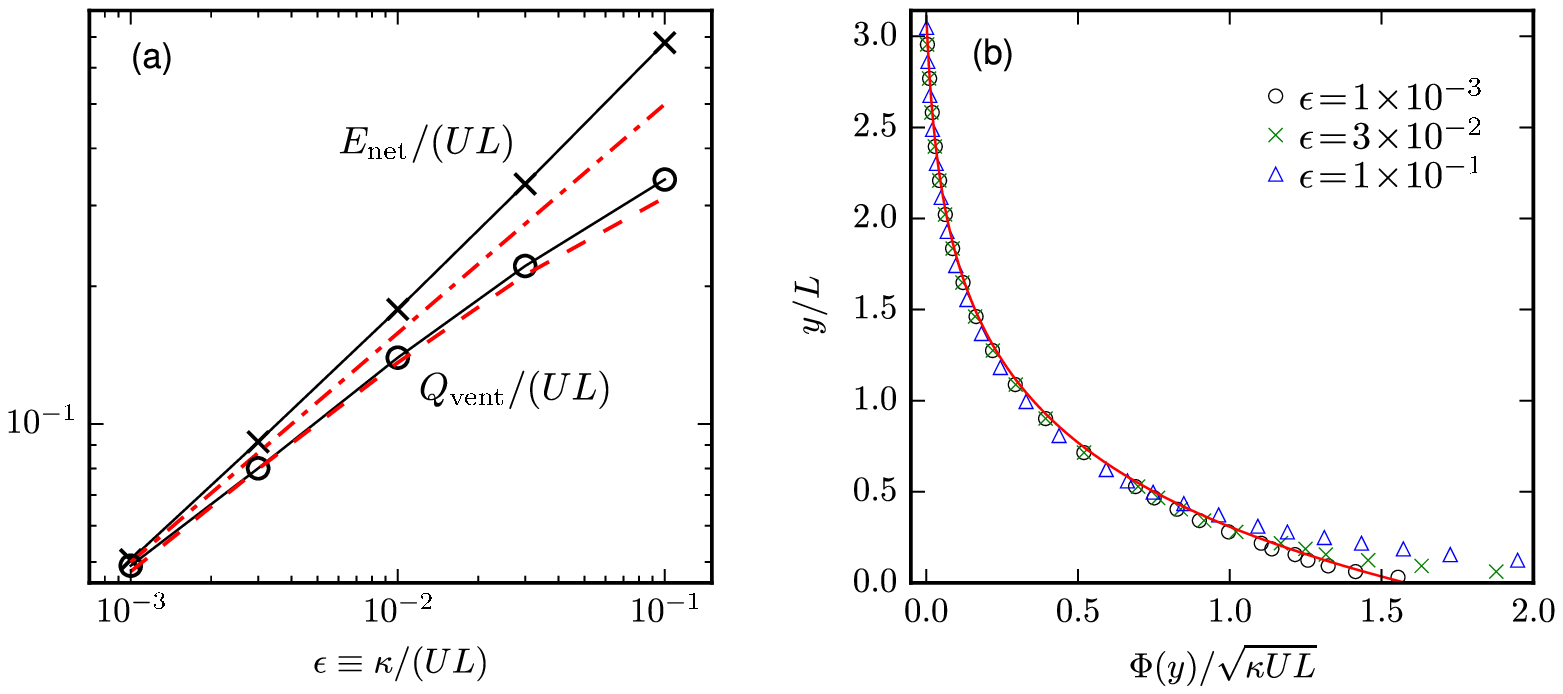}
\caption{(a) Surface evaporation rate $\Enet$ and boundary-layer ventilation $\Qvent$ defined by \Eq{qvent}: the Monte Carlo estimates of $\Enet$ (crosses) and $\Qvent$ (circles) are compared with the asymptotic predictions $\Enet/(UL)\propto \eprt$ in \Eq{enet_th} (dash-dotted line) and $\Phi_{\rm FT}(\eprt L)$ from \Eq{phift} (dashed line). (b) Vertical moisture flux $\Phi(y)$: Monte Carlo results (symbols) are compared with the asymptotic prediction \Eq{phift}.}
\label{evap_flux}
\end{figure}

We are also interested in the moisture flux outside the source boundary layer. In this ``free troposphere'' of the model, \Eq{phiy} is dominated by the first term. Using $P_0$ for the cell interior \Eq{ptop} and for the condensation boundary layer \Eq{bl2} in \Eq{phiy}, we find (see \ref{app:phiy} for details)
\begin{equation}
\Phi_{\rm FT}(y) = \sqrt{\kappa UL}\sqrt{\frac{8}{\pi}}\,[q_s(y)-\qmin].
\label{phift}
\end{equation}
\FF{evap_flux}(b) plots the scaled $\Phi(y)$ from several Monte-Carlo simulations (see also \ref{app:num}) with different $\epsilon$ together with the prediction $\Phi_{\rm FT}(y)$. The collapse of all the data onto the theoretical curve for $y \gg \eprt L$ verifies the prediction. The position where the numerical results start to deviate from the theory indicates the thickness of the source boundary layer is of order $\eprt L$.

The value of the moisture flux at the top of the planetary boundary layer $\Qvent$ is of particular importance for atmospheric moisture transport as it represents the amount of moisture ventilated from the boundary layer \cite{Boutle10}. \FF{evap_flux}(a) demonstrates the good agreement between
\begin{equation}
\Qvent \equiv \Phi(\eprt L)
\label{qvent}
\end{equation}
measured from simulations and the prediction $\Phi_{\rm FT}(\eprt L)$. Like $\Enet$, $\Qvent$ increases with $U$. In fact, $\Qvent \approx \Enet$ for small $\epsilon$ showing the large-scale circulation acts like a conveyor belt: air parcels enter the source boundary layer at one end and travel within the layer to the other end where they exit, carrying with them almost all the moisture they pick up from the surface source.

\subsection{Surface precipitation rate}

As wet parcels emerge from the source boundary layer and move upward into regions of low saturation $q_s(y)$, condensation occurs. We assume that all condensed moisture becomes precipitation. With $y$ interpreted as altitude and assuming precipitation falls vertically, we can consider the distribution $R(x)$ of surface precipitation rate. $R(x)$ measured from Monte-Carlo simulations (as described in \ref{app:num}) can have a significant contribution $R_{\rm BL}(x)$ from the frequent condensation near $y=0$ induced by the Brownian small-scale motion described below \Eq{phiy}. We generally find $R_{\rm BL}$ depends very weakly on $x$. So we instead consider the net surface precipitation rate
\begin{equation}
\Rnet(x) = R(x) - R_{\rm BL}(x).
\label{Rnet}
\end{equation}
\FF{rain_rh}(a) shows $\Rnet$ normalised by $U$ from simulations of different $\epsilon$. When the circulation strength increases, the precipitation rate increases and the distribution of precipitation becomes more localised around $x=0$ (when $\kappa$ is held fixed), in line with the boundary-layer thickness scales like $\eprt$.

\begin{figure}
\centering
\includegraphics[width=\textwidth]{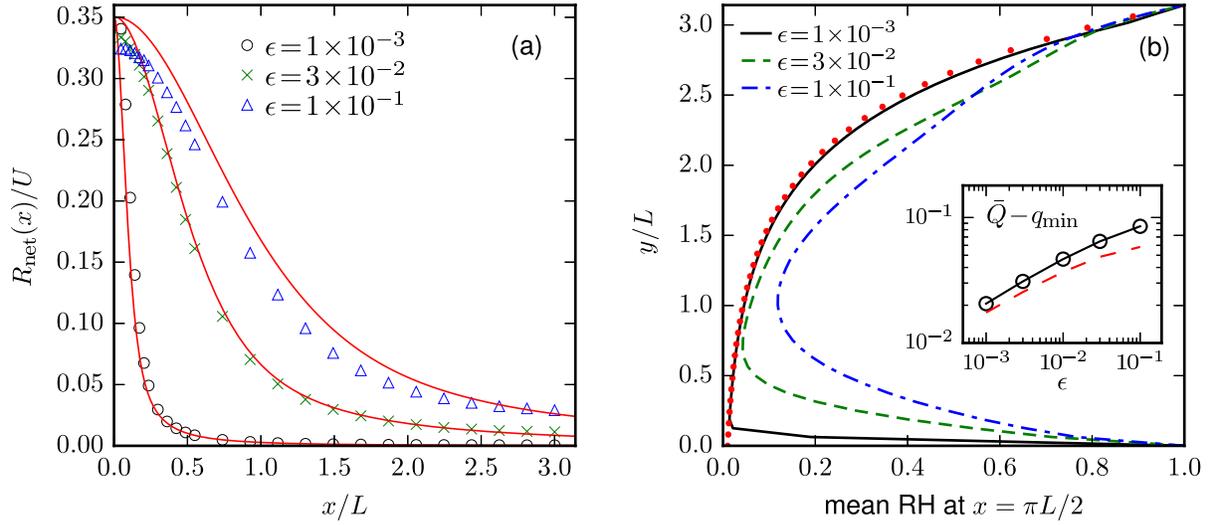}
\caption{(a) Distribution of net surface precipitation rate $\Rnet$ in \Eq{Rnet} for different $\epsilon\equiv\kappa/(UL)$. Simulation results (symbols) are compares with the prediction \Eq{rx} (solid lines). Precipitation increases and becomes more localised as the circulation strength $U$ increases. (b) Mean relative humidity profile at $x=\pi L/2$ for different $\epsilon$ compared with the limiting behaviour $\qmin/q_s(y)$ (dotted line). Inset: global specific humidity: the time-averaged $\bar Q$ of \Eq{qbardef} in the simulation (circles) is compared with the prediction \Eq{globalmoist} (dashed line).}
\label{rain_rh}
\end{figure}

We can calculate the leading-order $\Rnet(x)$ by considering transport inside the condensation boundary layer. Vertical transport is dominated by $v$, so we are in the ballistic limit studied by O'Gorman \& Schneider \cite{OGorman06}. Only parcels with $Q(t)=q_s[Y(t)]$ contribute to precipitation at time $t$. The amount of condensation from one such parcel between time $t$ and $t+\Delta t$ is
\begin{equation}
\Delta Q = q_s[Y(t) + v\Delta t] - q_s[Y(t)] \approx v\Delta t \dy{q_s}\bigg|_{Y(t)} < 0.
\end{equation}
The mean condensation rate at position $(x,y)$ is thus
\begin{equation}
\avg{\mathcal C} = -\lim_{\Delta t\rightarrow 0}\left\langle\frac{\Delta Q}{\Delta t}\right\rangle
= v(x,y)\dy{q_s}\int_{q_s(y)^-}^{q_s(y)^+}\! \tilde P_0(q|x,y)\, \dd q
\end{equation}
where the conditional PDF $\tilde P_0$ is obtained from \Eq{bl2}. We derive $\Rnet(x)$ by integration over $y$ to find
\begin{equation}
\Rnet(x) = \int_0^{\pi L}\! \avg{\mathcal C}\, \dd y 
= \alpha UL\qmax \int_0^\pi e^{-\alpha y} \sin y \bigg[1-\pi^2 G\bigg(\sqrt\frac{U}{\kappa L}x\sin y,3-\cos y\bigg)\bigg]\dd y
\label{rx}
\end{equation}
with $G$ given by \Eq{gg2}. Good agreement between this prediction and numerical results is seen in \Fig{rain_rh}(a).

\subsection{Relative humidity and global specific humidity}

When evaporation balances condensation, the system reaches a statistically steady state and the mean moisture distribution has a steady profile. The snapshot in \Fig{cell_qs} confirms that the relative humidity in the centre of the cell is near its minimum $\qmin/q_s(\pi L/2)$; the inset of \Fig{qcpdfbl}(c) shows that it decreases towards this minimum as $\epsilon\rightarrow 0$. Here we take a closer look by plotting in \Fig{rain_rh}(b) the mean relative humidity as a function of $y$ along a fixed $x=\pi L/2$. These profiles are obtained from Monte-Carlo simulations by averaging $Q/q_s(Y)$ over parcels as well as over time. \FF{rain_rh}(b) shows that the relative humidity decreases from  $y=0$, reaches a minimum, then increases with $y$, approximately as $\qmin/q_s(y)$. When the circulation strength $U$ increases with $\kappa$ fixed, the minimum relative humidity decreases and its location shifts toward $y=0$, or the direction of increasing $q_s$.

It is interesting to assess the dependence of the total water vapour content, as estimated by the global specific humidity $\bar Q$ in \Eq{qbardef}, on the circulation strength. The theoretical prediction for $\bar Q$ is given by the expectation value
\begin{equation}
\avg{\bar Q} = \iint\int_{\qmin}^{\qmax} q P(q,x,y) \, \dd q \dd x \dd y.
\end{equation}%
Clearly, $\avg{\bar Q }\to \qmin$ as $\epsilon \to 0$ since the area of the source and condensation boundary layers (where $Q \not= \qmin$) tends to $0$ with $\epsilon$.  The leading-order correction is controlled by the solution near the corner of the domain at $(x,y)=(0,0)$, where the moisture content is at its largest. A computation detailed in \ref{app:totalmoisture} gives
\begin{equation}
\avg{\bar Q} \sim \qmin - \epsilon^{1/2} \log \epsilon  \, \sqrt{\frac{2}{\pi^5}} (\qmax-\qmin),
\label{globalmoist}
\end{equation}
with the appearance of a logarithmic factor that can be traced to the streamline geometry near the corner. Thus, the total moisture increases with the diffusivity $\kappa$, that is, with the intensity of small-scale turbulence, and decreases as the strength of the large-scale circulation $U$ increases. The inset in \Fig{rain_rh}(b) confirms this result.

\section{Discussion and Conclusion}
\label{sec:conclude}

Motivated by the importance of synoptic-scale moisture transport in the atmosphere, we have studied two idealised problems based on the advection--condensation paradigm. The key element in both cases is that the advecting velocity has a large-scale coherent component in addition to small-scale white noise. The analytically tractable models introduced here capture some of the essential processes that control the large-scale dynamics of atmospheric water vapour, enabling us to examine the three-way interaction between large-scale advection, small-scale turbulence and moisture condensation.

We first study in \sect{sec:drying} the drying of a patch of initially saturated air and show how the action of a vortex speeds up the process. We predict the long-time decay of total moisture from the statistics of maximum excursion. The drying mechanism in this initial-value problem is responsible for the creation of a dry zone in the steady-state problem discussed in \sect{sec:forced}.

For the steady-state problem, we consider the single overturning cell \Eq{stream} on the $(x,y)$-plane with a moisture source at the boundary $y=0$. This can be interpreted as a large-scale circulating flow on an isentropic surface if we take $x$ and $y$ as the zonal and meridional directions respectively. Alternatively, this setup is a crude representation of the Hadley cell if we interpret $x$ as latitude and $y$ as altitude. This simple model produces some interesting features reminiscent of the atmosphere. First, a boundary layer near the source is formed as a result of the balance between large-scale and small-scale motions. This layer roughly mimics the atmospheric boundary layer whose role in moisture transport has  been investigated using idealised simulations with full physics \cite{Boutle10}. There is another boundary layer along the rising arm of the cell near $x=0$ -- the tropics of the model -- where  intense precipitation occurs. Second, we find that the moisture distributions inside both boundary layers are bimodal. The dry peak is a consequence of the subsidence of parcels with low humidity originated from the top of the cell. Satellite measurements indeed show the PDF of relative humidity over the whole tropics is bimodal although the PDF within a subregion could be unimodal \cite{Ryoo09}. Finally, the coherent stirring in the model produces a region of low relative humidity about the centre of the cell. A similar dry area in the subtropics is observed in the zonal-temporal-averaged relative humidity obtained from satellite measurements \cite{Sherwood10} and reanalysis data \cite{OGorman11}. The importance of these subtropical dry zones lies in their large influence on the radiation budget \cite{Pierrehumbert07} and the high sensitivity of such influence to water vapour feedback \cite{Held00}. Using an idealised model, O'Gorman at el. \cite{OGorman11} show a strong correlation between the position of the relative humidity minimum  and the inflection point of the saturation profile. In our model, the minimum is located at the edge of the source boundary layer, at an altitude of a few times $\sqrt{\kappa  L/U}$, independent of the details of the saturation profile.

There is a continuous interest in how climatological and seasonal variations in the strength and width of the Hadley cell \cite{Mitas05,Stachnik11} affect rainfall patterns. Some analysis associates the increase in tropical precipitation to the intensification of the Hadley circulation \cite{Quan04}. Increasing the strength of the circulation $U$ in our model does increase the amount of moisture injected into the system through surface evaporation, with the specific scaling $\sqrt{U}$ in the limit of strong circulation. This is balanced by a larger moisture flux and higher precipitation rate. The precipitation becomes more concentrated around $x=0$, with an extent that scales like $1/\sqrt{U}$; as a result the local precipitation intensity increases like $U$. An increase in circulation strength also leads to a drier atmosphere with humidity values that are only substantially larger than $\qmin$ in the increasingly small source and condensation boundary layers. Interestingly, the net moisture input \Eq{enet_th}, and as a consequence the total condensation above the source boundary layer, and the total moisture \Eq{globalmoist} depend only on $\qmin$ and $\qmax$ rather than on the full saturation profile $q_s(y)$ which only affects the spatial distribution of rainfall. It is known that changes in global-mean evaporation and precipitation with surface temperature are strongly constrained energetically \cite{Schneider10}. How well our simple qualitative conclusions apply to more complete models of the atmosphere remains to be assessed.

Previous work using simplified one-dimensional models has established the Lagrangian formulation of the advection--condensation paradigm as a promising strategy to investigate atmospheric water vapour. The present study provides a step forward in this direction through the analysis of a stochastic Lagrangian model that incorporates the dynamics of a two-dimensional large-scale circulation. An important extension in the future is to include the effects of latent heat by making temperature a dynamical variable and the saturation profile temperature-dependent. As it is often difficult to untangle the many interacting processes in full general circulation model simulations, idealised models such as the one introduced here can help to reveal the role of specific processes in controlling the distribution of water vapour in the atmosphere.
\medskip

\noindent
This work was supported by the UK EPSRC (Grant No.\ EP/I028072/1). YKT was partially supported by a Feasibility Grant from the EPSRC network Research on Changes of Variability and Environmental Risk (ReCoVER). We thank Darryn Waugh and Bill Young for many helpful discussions. Darryn's hospitality during YKT's short visit to Baltimore is very much appreciated.

\appendix

\section{Long-time decay of global specific humidity}
\label{larget}

Here we derive the long-time behavior of the expectation of the global specific humidity $\avg{\bar Q}$. We consider only the first term in the series \Eq{qbar2}. With $\hat t=\alpha^2\kappa t$ and introducing $\hat w=-\ln\hat q$, we have
\begin{equation}
\avg{\bar Q} \approx \frac{4\qmax}{\alpha^2R^2}\int_0^\infty\left[\frac{2\hat t}{\hat w}
+ \frac{J_2(z_1)}{z_1J_1(z_1)}\hat w \right]\exp\!\left(-\frac{z_1^2\hat t}{\hat w^2}-\hat w\right) \dd\hat w \,.
\end{equation}
The argument of the exponential function has a movable maximum at $\hat w_* = (2z_1^2\hat t)^{-1/3}$. Hence let $w=\hat w_*\hat w$ to obtain
\begin{align}
\avg{\bar Q} \approx \frac{4\qmax}{\alpha^2R^2}\int_0^\infty
\left[\frac{1}{w}+\frac{z_1^{1/3}J_2(z_1)}{J_1(z_1)}\frac{w}{(2\hat t)^{1/3}} \right] 
\exp\!\left[-(2z_1^2\hat t)^{1/3}\left(\frac{1}{2w^2}+ w\right)\right] \dd w \,.
\label{a2}
\end{align}
Applying Laplace's method to the integral in \Eq{a2} for $\hat t\gg 1$ leads to
\begin{align}
\avg{\bar Q} \sim \sqrt{\frac{2\pi}{3}}
\left[\frac{1}{z_1^{1/3}} + \frac{1}{(2\hat t)^{1/3}}\frac{J_2(z_1)}{J_1(z_1)}\right] 
(2\hat t)^{5/6}\exp\!\left[-\frac{3}{2^{2/3}}(z_1^2\hat t)^{1/3}\right] \,,
\end{align}
from which \Eq{asym} follows.

\section{Leading-order vertical moisture flux}
\label{app:phiy}
Working in dimensionless variables, we derive \Eq{phift} by computing the first term in \Eq{phiy} as follows. Let $\eprt\ll\delta_*\ll 1$. For $0<x<\delta_*$, we use \Eq{bl2} with $G$ given by \Eq{gg2} and $v\approx\sin y$ to obtain
\begin{align}
\int_0^{\delta_*} v\avg{Q}\,\dd x
&= \int_0^{\delta_*}\sin y \int_{\qmin}^{q_s} q\left\{(1-G)\big[\delta(q-q_s)-\delta(q-\qmin)\big]+\delta(q-\qmin)\right\}\, \dd q\dd x \nonumber \\
&= \eprt[q_s(y)-\qmin]\int_0^\infty[1-G(\hat\psi,\gamma)]\,\dd \hat\psi + \qmin\delta_*\sin y.
\label{b1}
\end{align}
Here, we have used the transformation $\hat\psi=\epsilon^{-1/2}x\sin y$ and taken the limit $\epsilon\rightarrow 0, \delta_*/\eprt\rightarrow\infty$. Integrating \Eq{heat} over all $\hat\psi$ and noting that $\partial_{\hat\psi}G=0$ at $\hat\psi=0,\infty$ shows that the $\hat\psi$-integral in \Eq{b1} is independent of $\gamma$ and hence can be evaluated by replacing $G(\hat\psi,\gamma)$ with $G(\hat\psi,2)$ to give $\sqrt{8/\pi}$.
Next, for $\delta_*<x<\pi$, using \Eq{ptop} and $v=\cos x\sin y$, we find
\begin{equation}
\int_{\delta_*}^\pi v\avg{Q}\,\dd x = \int_{\delta_*}^\pi\!\cos x\sin y\int_{\qmin}^{q_s} q\, \delta(q-\qmin)\,\dd q\dd x
= -\qmin\delta_*\sin y \quad\text{as}\quad \delta_*\rightarrow 0.
\label{b2}
\end{equation}
Combining the results in \Eq{b1} and \Eq{b2} and reverting to dimensional variables gives \Eq{phift}.

\section{Monte-Carlo simulation diagnostics}
\label{app:num}

In a Monte-Carlo simulation with $N$ parcels, let $N_p(y,t)$ be the number of parcels crossing a given height $y$ in either direction between time $t$ and $t+\Delta t$. Assuming that all parcels have the same total air mass $M$, the $i$th parcel carries a moisture mass of $Q_iM$. Recall that $\Phi(y)$ in \Eq{phiy} is the rate of upward transport of moisture mass across $y$ divided by the mass density $NM/(\pi L)^2$. With $\sigma_i(t)$ the sign of $\dd Y_i/\dd t$, we estimate $\Phi(y,t)$ from simulation data by summing over these set of $N_p$ parcels as follows:
\begin{equation}
\Phi(y,t)
=\frac{(\pi L)^2}{N\Delta t} \sum_{i=1}^{N_p} \sigma_i\, q^{\!*}_i(t)
\quad\text{where}\quad
q^{\!*}_i(t) = 
\begin{cases}
\min[Q_i(t),q_s(y)] & \text{if } \sigma_i >0, \\
Q_i(t) & \text{if } \sigma_i < 0.
\end{cases}
\end{equation}
The statistically steady $\Phi(y)$ is then obtained by averaging $\Phi(y,t)$ over $t$.

To estimate the distribution $R(x)$ of surface precipitation rate, we divide the surface into $N_b$ bins of width $\Delta x = \pi L/N_b$. Denoted by $N_r(x,t)$ the number of parcels that undergo condensation at time $t$ and whose positions $X_i(t)$ fall in $(x-\Delta x/2,x+\Delta x/2]$. Summation over this set of parcels gives the total mass of precipitation per unit time about $x$ which defined $R(x,t)$:
\begin{equation}
\frac{NM}{(\pi L)^2}R(x,t)\Delta x \equiv \frac{1}{\Delta t}\sum_{i=1}^{N_r}[Q_i-q_s(Y_i)]M.
\end{equation}
It follows that
\begin{equation}
R(x,t) = \frac{\pi L}{\Delta t}\frac{N_b}{N} \sum_{i=1}^{N_r}[Q_i-q_s(Y_i)].
\end{equation}
We then average $R(x,t)$ over $t$ to get $R(x)$. $R_{\rm BL}(x)$ in \Eq{Rnet} is calculated similarly except that only parcels inside the source boundary layer are included in the summation.

\section{Global specific humidity}
\label{app:totalmoisture}

The difference between 
\begin{equation}
\avg{\bar Q} = \frac{1}{\pi^2} \iint \avg{Q} \, \dd x \dd y = \iint \int_{\qmin}^{\qmax} q P(q,x,y) \, \dd q \dd x \dd y
\end{equation}
(in dimensionless variables) and $\qmin$ arises from the source and condensation boundary layers. Asymptotically, most of the area of this region is located near the corner where $x,\,y \ll 1$ and 
\begin{equation}
\avg{Q} - \qmin \sim (\qmax-\qmin) \erfc \left(\frac{x y}{2 \sqrt{2} \epsilon^{1/2}} \right)
\label{Qqmin}
\end{equation}
according to \Eq{bl1} and \Eq{bl2}.
We now pick $\epsilon^{1/4} \ll \delta_* \ll 1$ and integrate \Eq{Qqmin} for  $(x,y) \in [0,\delta_*]^2$ to find
\begin{align}
\int_0^{\delta_*} \int_0^{\delta_*} \left(\avg{Q} - \qmin \right) \, \dd x \dd y &= {\epsilon^{1/2}(\qmax-\qmin)}
\int_0^{\delta_*^2/\epsilon^{1/2}} \left[ \frac{2\sqrt{2}}{{\sqrt{\pi}}} \, \frac{1-e^{-\xi^2/8}}{\xi}  + \erfc\left(\frac{\xi}{2\sqrt{2}} \right)\right] \, \dd \xi \nonumber \\ 
&= \frac{2 \sqrt{2} \epsilon^{1/2} {(\qmax-\qmin)}}{\sqrt{\pi}} \left[ \log \left(\frac{\delta_*^2}{\epsilon^{1/2}}\right) + O(1) \right],
\label{aaa}
\end{align}
where we have defined $\xi=\delta_* x /\epsilon^{1/2}$. Ignoring the term in $\log \delta_*^2$, this gives the result \Eq{globalmoist} for the global specific humidity. This term in fact cancels out when we account for the rest of the source and condensation boundary layers, that is, for the regions $(x,y) \in [\delta_*,\pi] \times [0,\mu]$ and $(x,y) \in [0, \mu] \times [\delta_*,\pi]$ for some $\epsilon^{1/2} \ll \mu \ll 1$. For the source boundary layer, we have from \Eq{bl1}
\begin{align}
&\int_0^\mu \int_{\delta_*}^\pi \left(\avg{Q}-\qmin\right) \,  \dd x \dd y \sim {(\qmax-\qmin)} \int_{\delta_*}^\pi \dd x \int_0^\mu \erfc \left( \frac{y \sqrt{1-\cos x}}{2 \epsilon^{1/2}} \right) \, \dd y \nonumber \\ 
\sim& \frac{2 \sqrt{2} \epsilon^{1/2}{(\qmax-\qmin)}}{\sqrt{\pi}} \tanh^{-1}\left(\cos\frac{\delta_*}{2}\right) = \frac{2 \sqrt{2} \epsilon^{1/2}{(\qmax-\qmin)}}{\sqrt{\pi}} \left[ - \log \delta_* + O(1) \right].
\label{sourceBL}
\end{align}
The contribution of the condensation boundary layer is more complicated because of the variable $q_s(y)$ in \Eq{bl2} and the integral term in $G$ in \Eq{gg2}. However, this contribution is identical to \Eq{sourceBL} to the leading order because this is controlled by the limit $y \to 0$ of $P(q,x,y)$ for which $q_s(y) \to \qmax$ and the integral term vanishes. 
Together the two contributions cancel the $\log \delta^2_*$ term in \Eq{aaa} as claimed.

\bibliographystyle{mybst}

\end{document}